\documentclass[aps,twocolumn,showpacs]{revtex4}
\usepackage{dcolumn}
\usepackage{graphicx}
\usepackage{amsmath}
\usepackage{amsfonts}
\usepackage{amssymb}
\usepackage{psfrag}
\usepackage{wrapfig}
\usepackage{subfigure}
\usepackage{makeidx}
\usepackage{bm}
\usepackage{epsf}

\begin{document}

\title{State Transition Induced by Higher-order Effects and Background Frequency}
\author{Chong Liu$^1$}
\author{Zhan-Ying Yang$^1$}\email{zyyang@nwu.edu.cn}
\author{Li-Chen Zhao$^{1}$}\email{zhaolichen3@163.com}
\author{Wen-Li Yang$^2$}
\address{$^1$School of Physics, Northwest University, Xi'an
710069, China}
\address{$^2$Institute of Modern Physics, Northwest University, Xi'an
710069, China}
\date{November 05, 2014}
\begin{abstract}
State transition between the Peregrine rogue wave and w-shaped traveling wave induced by higher-order effects and background frequency is studied.  We find that this intriguing transition, described by an exact explicit rational solution, is consistent with the modulation instability (MI) analysis that involves MI region and stability region in low perturbation frequency region. In particular, the link between the MI growth rate and transition characteristic analytically demonstrates that, the size characteristic of transition is positively associated with the reciprocal of zero-frequency growth rate.
Further, we investigate the case for nonlinear interplay of multi-localized waves.
It is interesting that the interaction of second-order waves in stability region features a line structure, rather than an elastic interaction between two w-shaped traveling waves.
\end{abstract}
\maketitle
Recently, significant progresses have been made on the
experimental observation of the Peregrine rogue wave \cite{Peregrine} in nonlinear systems, including optical fiber \cite{Kibler}, water-wave
tanks \cite{water}, and plasmas \cite{p}.
It indicates that this wave, with features of a high amplitude and double localization \cite{rw1,rw2}, appears as a result of modulation
instability (MI) \cite{MI1,review} of weakly modulated plane wave. More specifically, rogue waves exist only in the MI subregion where the instability frequency band is close to the vanishing frequency \cite{Baronio}. In this regard, rogue wave has a consistent structure in the standard nonlinear Schr\"{o}dinger equation (NLSE) where the MI characteristic remains invariant in nature. Nevertheless, the MI often exhibits some interesting features when the additional physical effects are taken into consideration, such as cross-phase modulation \cite{Forest, Zhao}, higher-order perturbation terms \cite{Wright}, etc. Thus, it is important to study the rogue wave property induced by the features of MI growth rate distribution.

Recent studies demonstrate that rogue wave can exhibit structural diversity beyond the reach of the standard NLSE in presence of higher-order effects \cite{h0,h1,h2,h3,h4,h5,h6,h7,h8,h9,h10,h11}. However, to our knowledge, less attention has been paid to analyzing rogue wave property in combination with the distribution characteristic of corresponding MI growth rate.
Here we find that, with some higher-order perturbation terms (the third-order dispersion
and delayed nonlinear response term), the MI growth rate shows a non-uniform distribution characteristic in low perturbation frequency region, in particular, it opens up a stability region as background frequency changes (see Fig. 1). Thus we anticipate that there will be some interesting physical properties as a rogue wave evolves and approaches the stability region.

As a starting point, we address the problem via a completely integrable higher-order NLSE{---}the Hirota equation (HE) \cite{Hirota}{---}which involves the higher-order perturbation effects.
In dimensionless form, the HE reads
\begin{eqnarray}
iE_z+\frac{1}{2}E_{t t}+|E|^2E-i\beta(E_{t t t}+6|E|^2E_t)=0,
\end{eqnarray}
where $z$ is the propagation variable, $t$ is the retarded time in a moving frame with
the group velocity, and $E(t,z)$ is the slowly varying envelope of
the wave field. The real parameter $\beta$ is introduced to be responsible for the third-order dispersion
and delayed nonlinear response term, respectively. If $\beta=0$, it reduces to the standard NLSE.
The existence of rogue waves in the HE has been recently demonstrated in \cite{h1,h2,h3}. It was reported that the HE allows only a tilted rogue wave structure
arising from the higher-order effects.

In order to study the state transition in the HE, we first pay our attention to the standard linear stability analysis.
Here, we take the expression of background solution with a general form
\begin{equation}
E_{0}=a e^{i\theta}, \theta=q t+[a^2-q^2/2+\beta(6qa^2-q^3)]z,
\end{equation}
\noindent where $a$ and $q$ represent amplitude and frequency, respectively.
A perturbed nonlinear background can be written as
$E_{p}=[a+p]e^{i\theta}$,
where $p(t, z)$ is a small
perturbation
which satisfies a linear
equation. Generally, $p(t,z)$ is given by collecting the Fourier modes as
$p=f_+e^{i(Qt+\omega z)}+f_{-}^{*}e^{-i(Qt+\omega^* z)}$, where
$f_+$, $f_{-}^{*}$ are small amplitudes,
and the propagation parameters $Q$ and $\omega$ are assumed to be real and complex, respectively.
A substitution of the perturbed
solution $E_p$ into Eq. (1), followed by linearization and the process for the occurrence of
baseband MI \cite{Baronio}, yields the
dispersion relation
\begin{equation}
\omega=2(6a^2\beta-q-3q^2\beta-Q^2\beta)\pm\sqrt{(Q^2-4a^2)(1+6q\beta)^2}.
\end{equation}
If $\textrm{Im}\{\omega\}<0,$ MI exists in the perturbation frequency region $-2a<Q<2a$, 
\begin{figure}[htb]
\centering
\includegraphics[height=52mm,width=70mm]{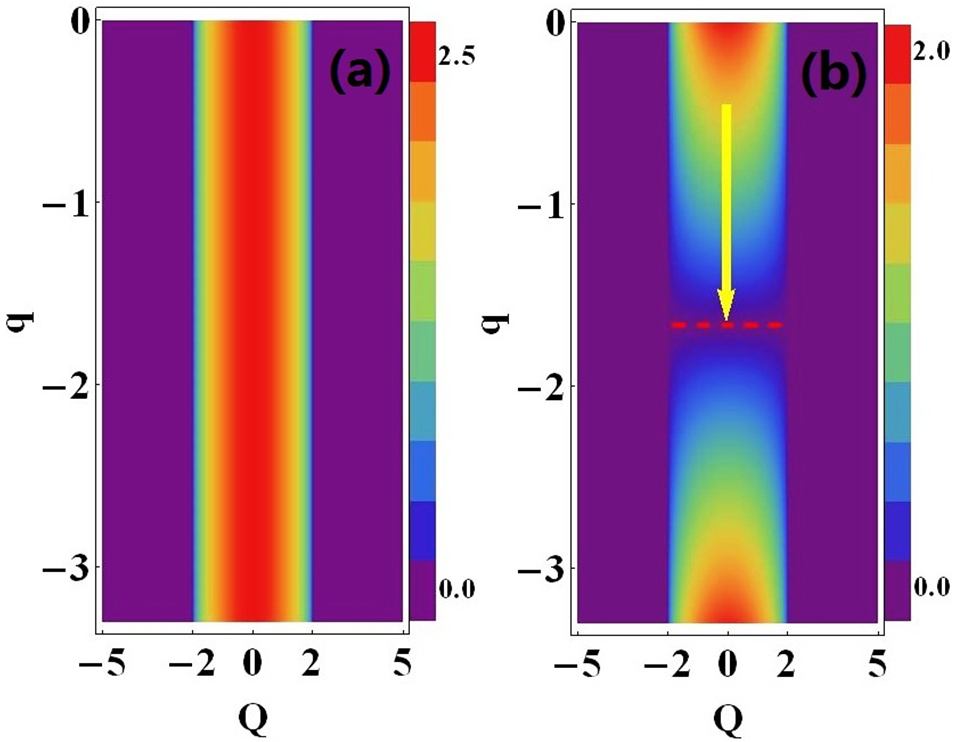}
\caption{Characteristics of MI growth rate $G$ on $(Q, q)$ plane with $a=1$, (a) without higher-order effects ($\beta=0$); (b) with higher-order effects ($\beta=0.1$). It shows obviously that the higher-order effects impact the MI growth rate distribution as $q$ changes.
Here the dashed red line represents the stability region in the perturbation frequency region $-2a<Q<2a$, which is
given as: $q=q_s=-1/(6\beta)$. An arrows in the zero-frequency MI subregion depicts rogue wave evolution process from the MI region to the stability line, i.e., $q\rightarrow q_s$.}
\end{figure}
\begin{figure}[htb]
\centering
\includegraphics[height=42mm,width=70mm]{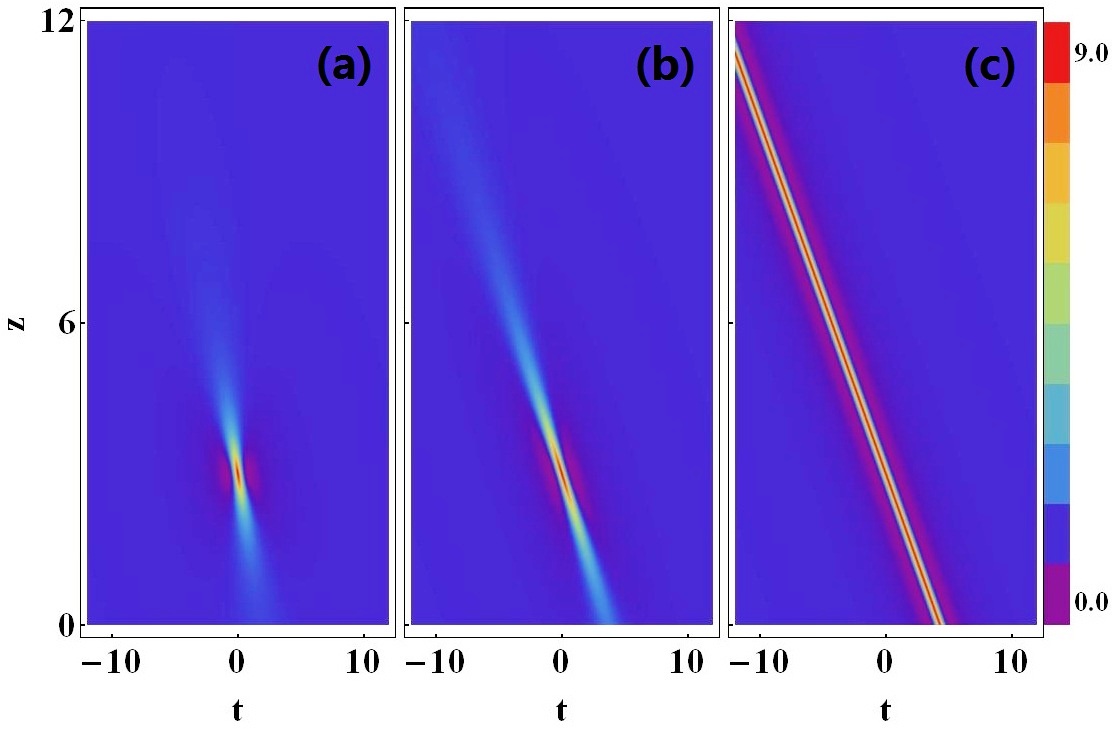}
\caption{Transition of first-order localized waves $|E_1(t,z)|^2$ from rogue wave to w-shaped traveling wave: (a) $q=0$, (b) $q=q_s/2$, (c) $q=q_s$, all obtained under $a=1$, $\beta=0.1$, $z_0=3$, and $t_0=0$. This corresponds to the process from MI region to stability line described by the arrows in Figure 1(b). Here (a) and (b) show the classical tilted rogue waves in the HE, while (c) shows a w-shaped traveling wave.}
\end{figure}
and its growth rate is defined as: $G=-\textrm{Im}\{\omega\}$.
Indeed, if $G>0$, perturbations grow exponentially
like $\exp(Gz)$ at the expense of pump waves.

In Fig. 1, we present the characteristics of MI on $(Q, q)$ plane. In the case $\beta=0$, i.e., the standard NLSE, it shows in Fig.1(a) that, the MI exists in the region $-2a<Q<2a$, and the distribution of the zero-frequency MI growth rate is uniform. However in the case $\beta\neq0$, the MI of the HE shows a non-uniform distribution characteristic as background frequency $q$ changes. As shown in Fig. 1(b), the MI growth rate distribution is symmetric about the $q=q_s=-1/(6\beta)$ line (red dashed line) where the corresponding growth rate is vanishing in the low perturbation frequency region. Moreover, the larger that $|q-q_s|$ is, the higher that $G$ becomes. Specifically, the zero-frequency MI growth rate can be written from Eq. (3) as: $G_0=2a|\frac{q-q_s}{q_s}|$. Since rogue wave exists in the zero-frequency MI subregion \cite{Baronio}, it will be interesting to study rogue wave property induced by the uneven distribution of zero-frequency MI growth rate.

To this end, we will turn our attention to the exact rational solutions on the background (2).
These solutions are constructed
by means of the Darboux transformation method, and their concise and general forms can be expressed as:
\begin{equation}
E_1(t,z)=E_{0}\left[\frac{4+8ia^2(1-q/q_s)\xi}{1+4a^4(1-q/q_s)^2\xi^2+4a^2(\tau-\upsilon\xi)^2}-1\right],
\end{equation}
where $\upsilon=q+(2a^2-q^2)/(2q_s)$, $\xi=z-z_0$, $\tau=t-t_0$, $z_0$ and $t_0$ are arbitrary real number which determine
the center of solution.
Interestingly, expression (4) consists of two types of localized waves with different dynamics properties, depending on the specific choice of the value $q$. Specifically, in the case $q\neq q_s$, it features double localization
and possesses a single peak and two zero-amplitude points, which has been identified as the prototype of the rogue wave phenomena [see Figs. 2(a) and 2(b)]. Indeed, if $a=1$, $q=0$, solution (2) reduces to the simplest rogue wave expression form in the HE \cite{h1}. While in the case $q=q_s$, the expression represents a traveling wave solution corresponding to the
stability region, and reads
\begin{equation}
E_{1s}(t,z)=ae^{i\theta_s}\left[\frac{4}{1+4a^2(\tau-\upsilon_s\xi)^2}-1\right],
\end{equation}
where $\theta_s=q_s\tau-q_s^2\xi/3$, and $\upsilon_s=(2a^2+q_s^2)/(2q_s)$. This wave features a soliton-like structure
with a stable peak $|E_{1s}|_p=3a$, and two stable valleys $|E_{1s}|_v=0$, which can be called as w-shaped traveling wave \cite{w}. It is worth noting that such unique solution (5) is specific to the HE since it cannot exist in the standard NLSE.

Figure 2 shows the transition characteristics of the wave evolution from the MI region ($q=0$) to the stability line ($q=q_s$) corresponding to the MI growth rate arrows in Fig. 1. As $q\rightarrow q_s$, the rogue wave becomes more elongated [Figs. 2(a) and 2(b)], which is corresponding to the smaller growth rate. When $q=q_s$, the rogue wave is translated into the w-shaped wave [Fig. 2(c)] with the vanishing growth rate.
In order to better understand the transition property,
we then define the aspect ratio of localized waves to describe their localization characteristics based on Eq. (4).
Its expression is given as :$\frac{\Delta L}{\Delta W}=\frac{1}{\sqrt{3}a}|\frac{q_s}{q-q_s}|$, where the length $\Delta L$ is defined by the half-value distance of the peak evolution on the background, and the width $\Delta W$ stands for the distance between the two zero-amplitude points. As shown in Fig. 3, the curve (solid line) of the aspect ratio is symmetric about $q=q_s$ line. As $q\rightarrow q_s$, the value of $\frac{\Delta L}{\Delta W}$ increases rapidly, and eventually becomes infinity when $q=q_s$. Namely, the emergence of w-shaped wave is associated with delocalization of propagation direction in presence of the higher-order effects.

We highlight that, the interesting finding is that the size characteristic of the waves turns out to be positively associated with the reciprocal of the MI zero-frequency growth rate, i.e., $\frac{\Delta L}{\Delta W}\sim \frac{1}{G_0}$, as shown in Fig. 3. Therefore, the link between the transition characteristic and the non-uniform distribution of MI growth rate is established.

\begin{figure}[htb]
\centering
\includegraphics[height=45mm,width=70mm]{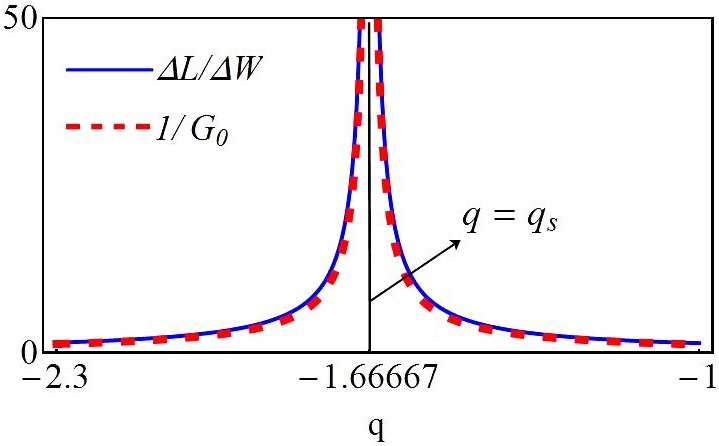}
\caption{Characteristics of the wave size $\Delta L/\Delta W$ (blue solid line) and the reciprocal of the zero-frequency growth rate of modulation instability $1/G_0$ (red dashed line).}
\end{figure}
\begin{figure}[htb]
\centering
\includegraphics[height=42mm,width=70mm]{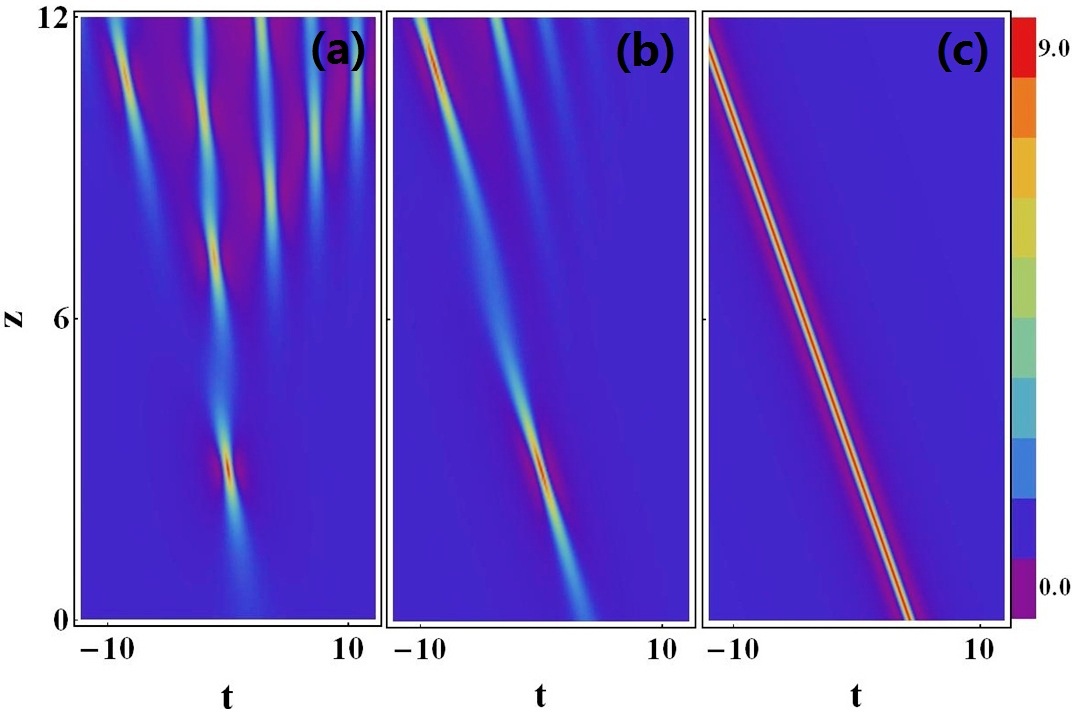}
\caption{Numerical simulations for the transition of first-order localized waves in Fig. 2.}
\end{figure}
\begin{figure}[htb]
\centering
\includegraphics[height=42mm,width=70mm]{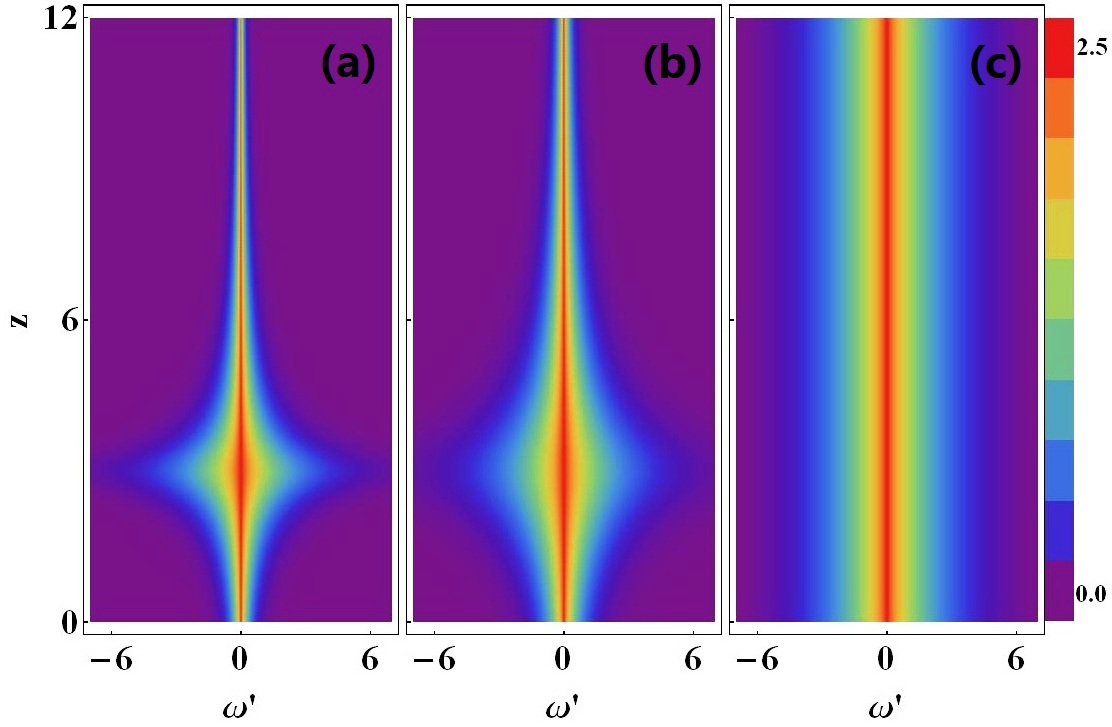}
\caption{Spectral dynamics $|F(\omega,z)|$ for the transition in Fig. 2. Here (a) and (b) show the typical rogue wave spectrum that gets extremely broadened at maximally-compressed peak ($z_0=3$). In contrast, (c) shows a stable broad spectrum structure of the w-shaped traveling wave.}
\end{figure}
\begin{figure}[htb]
\centering
\includegraphics[height=42mm,width=70mm]{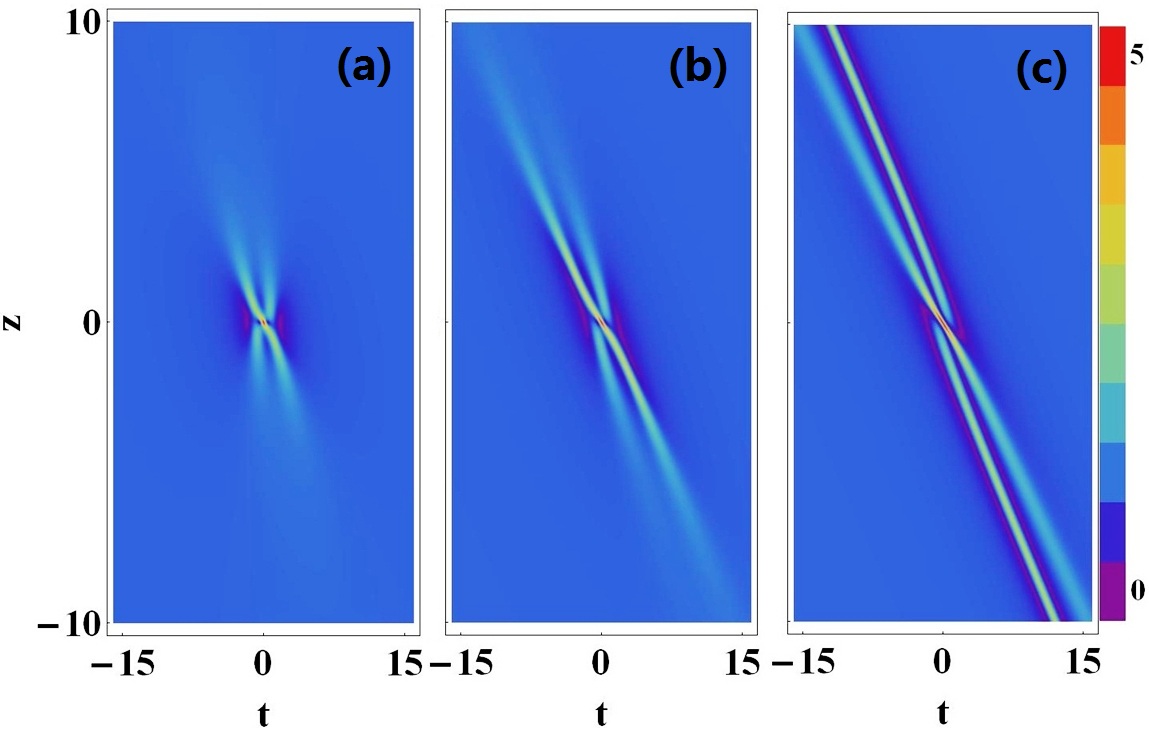}
\caption{Transition of second-order waves $|E(t,z)|$ with (a) $q=0$, (b) $q=q_s/2$, (c) $q=q_s$, all obtained under $a=1$, $\beta=0.1$, $z_0=0$, and $t_0=0$. (a) and (b) show the second-order rogue waves with a single peak, and the maximum peak value is always 5 times of the background. (c) shows the local structure of localized wave interaction in stability region.}
\end{figure}
\begin{figure}[htb]
\centering
\includegraphics[height=40mm,width=70mm]{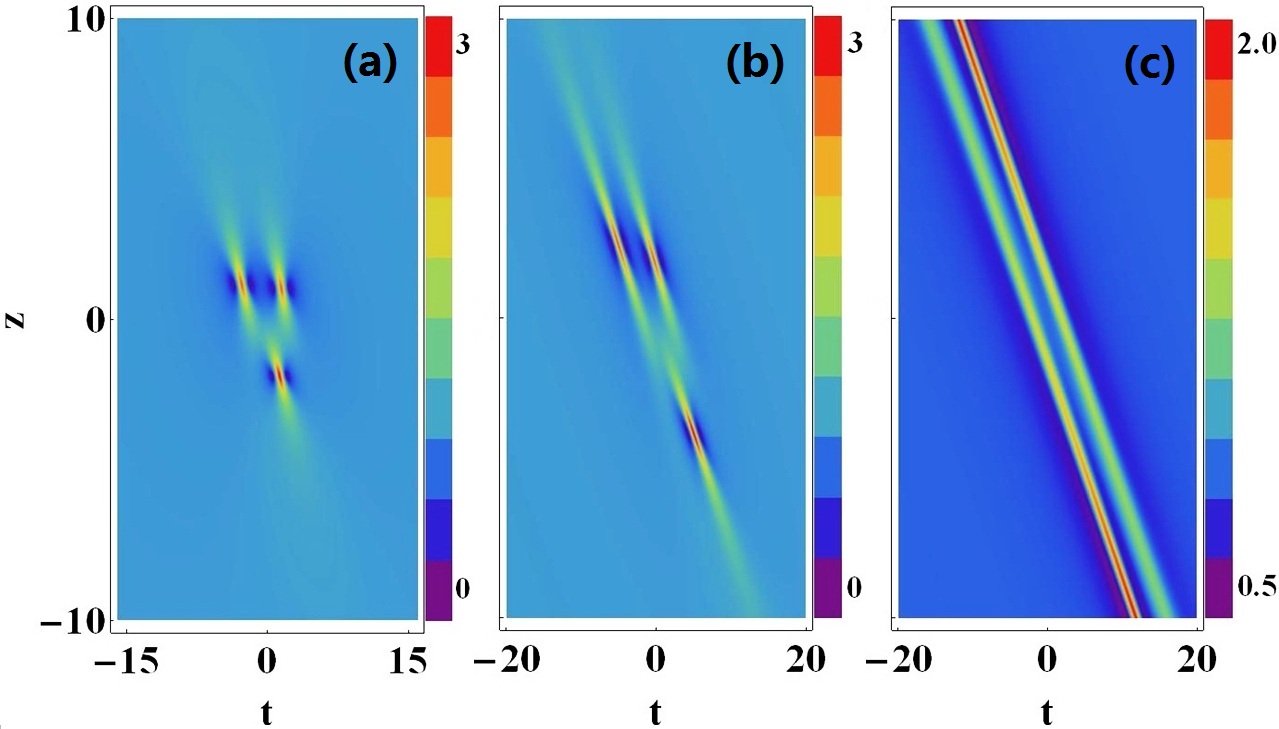}
\caption{Transition of second-order waves $|E(t,z)|$ with (a) $q=0$, (b) $q=q_s/2$, (c) $q=q_s$.
(a) and (b) show the rogue wave triplets, i.e., three
separate first-order rogue waves (their maximum peak values are all 3 times of the background). (c) shows the local structure of localized wave interaction in stability region.}
\end{figure}

We also perform direct numerical simulations of Eq. (1) to verify the robustness of transition. As shown in Fig. 4, the numerical results agree well with the exact case (illustrated in Fig. 2) for the first six propagation distances. After that, the rogue waves exhibit a breakup state [Figs. 4(a) and 4(b)], while the wave in the stability region keeps stable [Fig. 4(c)]. This results confirm numerically the characteristic of the transition between the unstable rogue wave with the stable w-shaped wave.

\begin{figure}[htb]
\centering
\includegraphics[height=62mm,width=70mm]{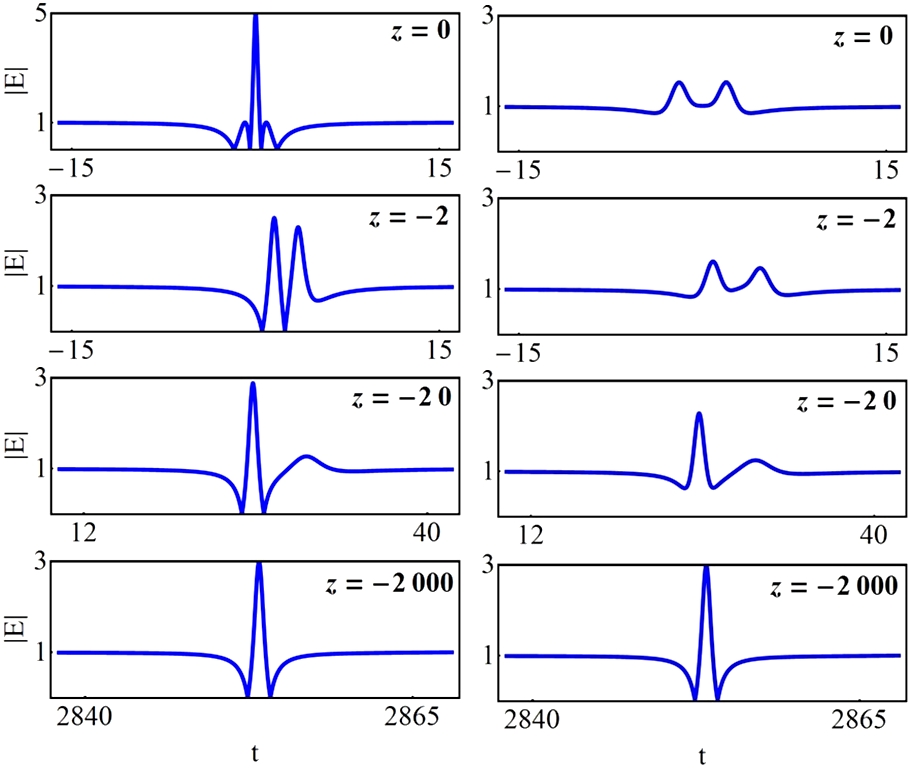}
\caption{The amplitude profile of wave interaction in stability region in Figs. 6(c) (left) and 7(c) (right) at $z=-2000,-20,-2,0$. Due to the interaction structure is symmetric about $(t_0,z_0)$, so that the case for $z>0$ is omitted.}
\end{figure}

Recently, it is demonstrated that the Peregrine rogue wave spectrum in the standard NLSE features a triangular shape and gets dramatically broadened at the maximally-compressed peak, which is corresponding to the onset of supercontinuum generation \cite{s2,s3,s4,s5}. Thus, it is meaningful to study the spectral property of the transition between the Peregrine rogue wave and the w-shaped traveling wave.
Here the analytic spectrum of solution (4) can be written with $a=1$ as:
\begin{eqnarray}
|F(\omega,z)|^2=2\pi\exp[-|\omega'|\sqrt{1+4(q-q_s)^2\xi^2/q_s^2}],
\end{eqnarray}
where $\omega'=\omega+q$. The corresponding spectrum dynamics is illustrated in Fig. 5.
We see that the spectrum of rogue wave [Figs. 5(a) and 5(b)] features a triangular structure and is greatly broadened at the maximally-compressed peak ($z_0=3$). This property coincides with the rogue-wave spectral dynamics reported in \cite{Kibler,s5}. As $q\rightarrow q_s$, the maximum value of spectrum width is unchanged, while the broadening velocity decreases gradually along the propagation direction $z$. When the transition occurs, the spectrum of the traveling wave becomes broad and stable [Fig. 5(c)]. This finding may suggest a way to manage a pulse to be a stable supercontinuum.

Finally, we would like to study the nonlinear interplay characteristic of the wave transition. It is well-known that the higher-order rogue wave exhibits rich structure characteristic \cite{t1,t2,t3,t4,t5,t6,t7}. Thus, it is necessary to answer how about the property of the higher-order wave transition.  As an example, we show the transition of second-order waves in Figs. 6 and 7. In the case $q\neq q_s$, the second-order wave describes a formation of the interaction between three first-order rogue waves. Figs. 6(a) and 7(a) exhibit two typical second-order rogue wave patterns, including single-peak structure and rogue wave triple \cite{rwt}. As $q\rightarrow q_s$,
the maximum peak value always remains the same while the aspect ratio changes [Figs. 6(b) and 7(b)], which is consistent with the case of first-order rogue wave.
In the case $q=q_s$, wave interaction
in the stability region [Figs. 6(c) and 7(c)] is obtained. Surprisingly, it shows the interaction with line structure, rather than an elastic interaction between two w-shaped traveling waves; specifically, for $z\rightarrow-\infty$, there is only one w-shaped traveling wave (see the corresponding amplitude profile in Fig. 8), once its amplitude decreases, a subwave with small amplitude is generated. As $z$ increases,
the two waves interplay with each other, and the amplitude of main wave decreases gradually while the amplitude of subwave increases. At the central position $(t_0,z_0)$, the two waves collide and form a large amplitude wave or a two-peak wave with small amplitude. They then separate after the collision, and undergo energy exchange. For $z\rightarrow\infty$, it restores the original shape of the first-order w-shaped traveling wave.
Here the excitation $|E(t,z)|^2$ above and below the background level are equal, since the conserved quantity of the HE, i.e., $\int_{-\infty}^{+\infty}\{|E(t,z)|^2-a^2\}d t=0$.

In summary, we have investigated the state transition of the localized waves on a plane-wave background arising from the higher-order effects in the HE. It shows that this transition is strictly related with the MI analysis that involves MI region and stability region. Numerical simulations demonstrate the robustness of the transition. In particular, we shed light on the relation between the MI growth rate and transition characteristic that, the aspect ratio characteristic of the waves turns out to be positively associated with the reciprocal of zero-frequency growth rate, i.e., $\frac{\Delta L}{\Delta W}\sim \frac{1}{G_0}$. This results will enrich our understanding of the relation between the MI characteristic and rogue-wave evolution property.
On the other hand, we find that the nonlinear wave interaction in stability region exhibits a line structure, rather than an elastic interaction between two w-shaped traveling waves. This interaction property is distinct from the previous results, including the elastic interaction between line rogue waves \cite{Ohta} induced in MI region, and the elastic collision between algebraic solitons \cite{Burde}. Similar studies could be extended to
other abundant higher-order NLSE systems where the MI could be affected by higher-order effects.

We are grateful to Professor Lu Li for his helpful discussions. This work has been supported by the National Natural Science Foundation of China (NSFC)(Grant No. 11475135), the ministry of education doctoral program funds (Grant No. 20126101110004), and NWU graduate student innovation funds (Grant No. YZZ13096).

\end{document}